\documentclass[showpacs,pra,preprint]{revtex4-1}
\usepackage{graphicx}
\usepackage{dcolumn}
\usepackage[latin5]{inputenc}
\usepackage{amssymb,amsmath,mathptmx,amsbsy,bm}
\def\bra#1{\mathinner{\langle{#1}|}}
\def\ket#1{\mathinner{|{#1}\rangle}}
\def\braket#1{\mathinner{\langle{#1}\rangle}}

\begin{document}
\title{Canonical operators and the optimal concentration of three-qubit Greenberger-Horne-Zeilinger states}
\author{Gokhan Torun}
\email{torung@itu.edu.tr}
\author{Ali Yildiz}
\email{yildizali2@itu.edu.tr}
\affiliation{Department of Physics, Istanbul Technical University, Maslak 34469, Istanbul, Turkey}
\date{\today}

\begin{abstract}
It is well known that quantum states that can be transformed into each other by local unitary transformations are equal from the information theoretic point of view. This defines equivalence classes of states and allows one to write any state with the minimal number of parameters  called the canonical form of the state.  We define the equivalence classes of local measurements such that local operations which transform states from one equivalence class into another with the same probability are equivalent. This equivalence relation allows one to write the operators with the minimal number of parameters, which we call canonical operators, and hence the use of the canonical operators simplifies the optimal manipulation of quantum states. We use the canonical local operators for the concentration of three-qubit Greenberger-Horne-Zeilinger states  and obtain the optimal concentration protocols  in terms of the unitary invariants of quantum states, namely, the bipartite concurrences and the three-tangle.
\end{abstract}

 \maketitle





\section{Introduction}

Quantum entanglement is not only the most fascinating aspect of quantum mechanics, but it is also used as a resource in quantum information processes. The use of quantum entanglement as a resource  requires a deep understanding of how the states are transformed under local operations and classical communication (LOCC). Two-qubit maximally entangled Einstein-Podolski-Rosen  (EPR) state, $\frac{1}{\sqrt{2}}(\ket{00}+\ket{11})$, is  used as a resource in perfect quantum information processes such as  teleportation \cite{teleportation1} and dense coding \cite{densecoding}. If the resource is a mixed state of two qubits, then the fidelity of the teleportation can be increased by the manipulation of the quantum state by LOCC \cite{eprdistill1,eprdistill2,eprdistill3,VerstaeteOptTele}. In the case where the resource is a nonmaximally entangled pure state, $\sqrt{1-a^2}\ket{00}+a\ket{11}$  $(0\leq a\leq 1/2)$, it is possible to  perform the task with some nonzero probability \cite{probabteleport1,probabteleport2}. An alternative way is concentrating the partial entanglement by local operations and then performing the information task with unit probability \cite{BennettConcent}. Optimal concentration of an EPR state can be obtained by a measurement of one of the qubits with the positive operator valued measurement (POVM)  elements
\begin{equation}\label{oper1a}
M_1=\frac{a}{\sqrt{1-a^2}}\ket{0}\bra{0}+\ket{1}\bra{1}, \quad
M_2=\frac{\sqrt{1-2a^2}}{\sqrt{1-a^2}}\ket{0}\bra{0}.
\end{equation}
This method is called the  Procrustean method \cite{BennettConcent}  as the state is either transformed to an EPR state with a probability of $P=2a^2$ or else to a separable state. It is also possible to obtain any partially entangled state from an EPR state with unit probability. One can obtain the  partially entangled state by the measurement of one of the qubits of an EPR state with POVM elements
\begin{equation}\label{oper2a}
M_1=\sqrt{1-a^2}\ket{0}\bra{0}+a\ket{1}\bra{1}, \quad
M_2=a\ket{0}\bra{0}+\sqrt{1-a^2}\ket{1}\bra{1}
\end{equation}
followed by local unitary transformations.

While the entanglement of two qubits  is well understood, many problems still remain unsolved for higher-dimensional systems. One of these problems is the optimal manipulation of states both from a single copy or from many copies. Optimal manipulation of a single copy of a state is of importance from the practical point of view,  as experimentalists are not able to perform joint operations on multiple copies of the system in general. Nielsen \cite{Nielsen1} used the algebraic theory of majorization and obtained necessary and sufficient conditions for the entanglement transformation of two pure states of a bipartite system. An optimal local conversion strategy of these states was proposed by Vidal \cite{Vidal1}.
Although  an extension of Nielsen's  theorem to three-qubit pure states has recently been obtained by Tajima \cite{tajima},
there is no straightforward generalization  to the multipartite systems.  In the three-qubit case, there are two classes of tripartite-entangled states which cannot be converted into each other by stochastic local operations and classical communication (SLOCC), namely, the Greenberger-Horne-Zeilinger (GHZ)  and  $W$ class states \cite{GHZ_W_inequiv,acin7}.
It has been shown that the minimal number of product terms for any given state remains unchanged under SLOCC and  the minimal number of product terms is three for $W$ class states.  Any GHZ class state, on the other hand,  can be written as the sum of two product vectors uniquely in the standard product form
\begin{equation}\label{GHZ2}
\ket{\psi_{\textrm{GHZ}}}=\sqrt{K}(c_{\delta}\ket{0}\ket{0}\ket{0}+s_{\delta}e^{i\varphi}\ket{\varphi_{A}}\ket{\varphi_{B}}\ket{\varphi_{C}}),
\end{equation}
where
\begin{eqnarray}
\ket{\varphi_{A}}&=&c_{\alpha}\ket{0}+s_{\alpha}\ket{1},\nonumber \\
\ket{\varphi_{B}}&=&c_{\beta}\ket{0}+s_{\beta}\ket{1}, \\
\ket{\varphi_{C}}&=&c_{\gamma}\ket{0}+s_{\gamma}\ket{1}, \nonumber
\end{eqnarray}
and $K=(1+2c_{\delta}s_{\delta}c_{\alpha}c_{\beta}c_{\gamma}c_{\varphi})^{-1}$ ($c_{\delta}$ and $s_{\delta}$ stand for $\cos{\delta}$ and $\sin{\delta}$, etc.). The ranges for the five parameters are  $\delta\in(0,\pi/4], \alpha,\beta,\gamma,\in(0,\pi/2]$, and $\varphi\in[0,2\pi)$.

The $W$  state, $\ket{W}=\frac{1}{\sqrt{3}}(\ket{001}+\ket{010}+\ket{100})$, and the GHZ  state
\begin{equation}\label{GHZ}
\ket{\textnormal{GHZ}}=\frac{1}{\sqrt{2}}(\ket{000}+\ket{111})
\end{equation}
are  the representatives of the  $W$ and GHZ classes, respectively. Any GHZ class state given by Eq. \eqref{GHZ2} can be obtained by the application of the invertible local operators
\begin{equation}\label{GHZop}
\ket{\psi_{\textrm{GHZ}}}=\frac{1}{\sqrt{p}}(A\otimes B\otimes C)\ket{\textnormal{GHZ}},
\end{equation}
where
\begin{eqnarray}\label{GHZop1}
A&=&c_{\delta}\ket{0}\bra{0}+s_{\delta}c_{\alpha}e^{i\varphi}\ket{0}\bra{1}+s_{\delta}s_{\alpha}e^{i\varphi}\ket{1}\bra{1},
\nonumber \\
B&=&\ket{0}\bra{0}+c_{\beta}\ket{0}\bra{1}+s_{\beta}\ket{1}\bra{1}, \\
C&=&\ket{0}\bra{0}+c_{\gamma}\ket{0}\bra{1}+s_{\gamma}\ket{1}\bra{1},\nonumber
\end{eqnarray}
and the probability of success, $p= \bra{\textnormal{GHZ}}A^{\dag}A\otimes B^{\dag}B\otimes C^{\dag}C\ket{\textnormal{GHZ}}$, is not necessarily the maximum probability that can be achieved. The optimal transformation of the GHZ state, given by Eq. \eqref{GHZ}, to any GHZ-type state, given by Eq. \eqref{GHZ2}, is still an open problem. Cui  {\it {et al.}} \cite{cui2} used the two-term product decomposition  of the GHZ class states to obtain the upper bounds for the optimal probability of transformation from a GHZ state to other states of the GHZ class. The complexity of the optimal transformation of multi-qubit systems is mainly due to the large numbers of parameters coming from the general local operators since the general local operators acting on qubits are two-by-two complex matrices in general.

The optimal manipulation of the GHZ and  $W$ states is important because they are used as a resource in quantum information processes \cite{Agrawal1,Gao,Dai,Wtele,Wtele2,zhang2}. The optimal distillation protocols for symmetric and asymmetric three-qubit $W$ states were proposed in \cite{yildiz}. Kintas and Turgut obtained an upper bound for the maximum probability of transforming
$N$-qubit $W$ states \cite{optimal3} and the necessary and sufficient conditions to obtain this upper bound  were presented by Cui {\it {et al.}} \cite{optimal1}. The optimal local transformations of flip and exchange symmetric multiqubit states were obtained in \cite{optimal2}. Sheng {\it {et al.}}  proposed practical two-step entanglement concentration protocols for some $W$ class states \cite{sheng}. The first GHZ state distillation protocols \cite{Optimal0} consisting of two stages, primary and secondary distillations, were not optimal.
Acin {\it {et al.}} \cite{Optimal_GHZ_distil} used one  successful branch protocol in the the optimal concentration problem starting with the two-term product decomposition of a GHZ class state and obtained some partial results.

We approach the quantum-state manipulation problem from a different perspective by using the canonical form of states presented in \cite{cano-form1,cano-form2} and the equivalence classes of local operators. The use of equivalence classes reduces the number of parameters and hence simplifies the manipulation problem since local operators  which transform the states from one equivalence class into another with the same probability are equivalent. Moreover, the conversion probabilities and measurement operators can be found in terms of the unitary invariants of the quantum state.

We present the paper as follows: In Sec. II, we start with the canonical form of a general GHZ class pure state and discuss the equivalence classes of local POVMs. These equivalence classes are used to obtain the maximum success probabilities and measurement operators for the optimal concentration of the GHZ states in Sec. III. In Sec. IV, we present discussion and summary.

\section{Canonical forms of states and operations on three-qubit pure states }
First, we define the notation  for the equivalence relations for states and operations. Two states $\ket{\phi}$ and $\ket{\phi'}$ are in the same equivalence class ($\ket{\phi}\sim\ket{\phi'}$)  if they can be transformed into each other by local unitary transformations. Two operators $A$ and $A'$ are in the same equivalence class ($A\equiv A'$) if they both transform states in one  equivalence class, $\ket{\psi_i}$, to states in some other equivalence class with the same probability of success, i.e.,  the operators $A$ and $A'$ are equivalent ($A\equiv A'$) if they satisfy the following relations:
\begin{equation}
\ket{\phi}=\frac{A \ket{\psi_i}}{\sqrt{\bra{\psi_i}A^{\dag}A\ket{\psi_i}}}\sim
\ket{\phi'}= \frac{A' \ket{\psi_i}}{\sqrt{\bra{\psi_i}(A')^{\dag}A'\ket{\psi_i}}},
P=\bra{\psi_i}A^{\dag}A\ket{\psi_i}=\bra{\psi_i}(A')^{\dag}A'\ket{\psi_i}, \forall \ \ket{\psi_i}.
\end{equation}

Following the approach presented by Acin {\it {et al.}} \cite{cano-form1,cano-form2}, we define the canonical form of the GHZ class states which we are going to use in the concentration. Any three-qubit state
\begin{equation}
\ket{\phi}=\sum_{ijk}{t_{ijk}\ket{ijk}}
\end{equation}
defines matrices  $T_0$ and $T_1$ by
\begin{equation}
\ket{\phi}=\sum_{jk}{T_{0,jk}\ket{0}\ket{jk}+T_{1,jk}\ket{1}\ket{jk}}.\nonumber
\end{equation}
Under the unitary transformation on the first qubit,   the matrices $T_0$ and $T_1$ transform as
\begin{eqnarray}
T_0'&=&u_{00}^AT_0+u_{01}^AT_1, \nonumber\\
T_1'&=&u_{10}^AT_0+u_{11}^AT_1 \quad , \quad
u_{tz}=\braket{t|U|z}.
\end{eqnarray}
It is always possible to make $\textnormal{det} T_0'=0$ and the unitary transformations on the second and third qubits diagonalize $T_0'$ which  bring the state to the  canonical form given by
\begin{equation}\label{canonform}
\ket{\psi}=\lambda_0\ket{000}+\lambda_1e^{i\varphi}\ket{100}+\lambda_2\ket{101}+\lambda_3\ket{110}+\lambda_4\ket{111},
\quad \lambda_i \geq 0.
\end{equation}

However, the parameters $\lambda_i$ and $\varphi$ do not uniquely determine the the state (the equivalence class) because there are two solutions for $\textnormal{det} T_0'=0$, and this leads to two sets of parameters for the canonical form. The state with the other set of parameters can be found to be
\begin{equation}\label{kanon}
\widetilde{\ket{\psi}}=\widetilde{\lambda_0}\ket{000}+\widetilde{\lambda_1}e^{i\widetilde{\varphi}}\ket{100}
+\widetilde{\lambda_2}\ket{101}
+\widetilde{\lambda_3}\ket{110}+\widetilde{\lambda_4}\ket{111},
\quad \widetilde{\lambda_i} \geq 0
\end{equation}
where
\begin{eqnarray}\label{dual1}
\widetilde{\lambda_0}&=&\lambda_0\sqrt{\frac{(\lambda_2^2+\lambda_4^2)(\lambda_3^2+\lambda_4^2)}{\lambda_0^2\lambda_4^2+\left|\lambda_2\lambda_3-e^{i\varphi}\lambda_1\lambda_4\right|^2}},\nonumber \\
\widetilde{\lambda_2}&=&\lambda_2\sqrt{\frac{\lambda_0^2\lambda_4^2+\left|\lambda_2\lambda_3-e^{i\varphi}\lambda_1\lambda_4\right|^2}{(\lambda_2^2+\lambda_4^2)(\lambda_3^2+\lambda_4^2)}},\nonumber \\
\widetilde{\lambda_3}&=&\lambda_3\sqrt{\frac{\lambda_0^2\lambda_4^2+\left|\lambda_2\lambda_3-e^{i\varphi}\lambda_1\lambda_4\right|^2}{(\lambda_2^2+\lambda_4^2)(\lambda_3^2+\lambda_4^2)}},\\
\widetilde{\lambda_4}&=&\lambda_4\sqrt{\frac{\lambda_0^2\lambda_4^2+\left|\lambda_2\lambda_3-e^{i\varphi}\lambda_1\lambda_4\right|^2}{(\lambda_2^2+\lambda_4^2)(\lambda_3^2+\lambda_4^2)}},\nonumber \\
\widetilde{\lambda_1}e^{i\widetilde{\varphi}}&=&-i\lambda_1\sin(\varphi)\sqrt{\frac{(\lambda_2^2+\lambda_4^2)(\lambda_3^2+\lambda_4^2)}{\lambda_0^2\lambda_4^2+\left|\lambda_2\lambda_3-e^{i\varphi}\lambda_1\lambda_4\right|^2}}+\nonumber \\
& &\frac{\lambda_1\cos(\varphi)(\lambda_4^2(\lambda_2^2+\lambda_3^2+\lambda_4^2)-\lambda_2^2\lambda_3^2)
+\lambda_2\lambda_3\lambda_4(\lambda_0^2+\lambda_1^2-\lambda_2^2-\lambda_3^2-\lambda_4^2)}
{\sqrt{(\lambda_2^2+\lambda_4^2)(\lambda_3^2+\lambda_4^2)(\lambda_0^2\lambda_4^2+\left|\lambda_2\lambda_3-e^{i\varphi}\lambda_1\lambda_4\right|^2)}}.\nonumber
\end{eqnarray}
The equivalence of the states given by Eqs. \eqref{canonform} and \eqref{kanon} under local unitary transformations can be explicitly shown as follows: the following unitary operators
\begin{eqnarray}
U_A=\left(
\begin{array}{cc}
 \frac{\lambda _2 \lambda _3-e^{i \varphi } \lambda _1 \lambda _4}{\sqrt{\left|\lambda _2 \lambda _3-e^{i \varphi } \lambda _1 \lambda _4\right|{}^2+\lambda _0^2 \lambda _4^2}} & \frac{\lambda _0 \lambda _4}{\sqrt{\left|\lambda _2 \lambda _3-e^{i \varphi } \lambda _1 \lambda _4\right|{}^2+\lambda _0^2 \lambda _4^2}} \\
 -\frac{\lambda _0 \lambda _4}{\sqrt{\left|\lambda _2 \lambda _3-e^{i \varphi } \lambda _1 \lambda _4\right|{}^2+\lambda _0^2 \lambda _4^2}} & \frac{\lambda _2 \lambda _3-e^{-i \varphi } \lambda _1 \lambda _4}{\sqrt{\left|\lambda _2 \lambda _3-e^{i \varphi } \lambda _1 \lambda _4\right|{}^2+\lambda _0^2 \lambda _4^2}}
\end{array}
\right),\\
U_B=\left(
\begin{array}{cc}
 \frac{\lambda _2}{\sqrt{\lambda _2^2+\lambda _4^2}} & \frac{\lambda _4}{\sqrt{\lambda _2^2+\lambda _4^2}} \\
 -\frac{\lambda _4}{\sqrt{\lambda _2^2+\lambda _4^2}} & \frac{\lambda _2}{\sqrt{\lambda _2^2+\lambda _4^2}}
\end{array}
\right),\quad
U_C=
\left(
\begin{array}{cc}
 \frac{\lambda _3}{\sqrt{\lambda _3^2+\lambda _4^2}} & \frac{\lambda _4}{\sqrt{\lambda _3^2+\lambda _4^2}} \\
 -\frac{\lambda _4}{\sqrt{\lambda _3^2+\lambda _4^2}} & \frac{\lambda _3}{\sqrt{\lambda _3^2+\lambda _4^2}}
\end{array}
\right)\nonumber
\end{eqnarray}
applied to parties A, B and C respectively will transform the state Eq. \eqref{canonform} into the state Eq. \eqref{kanon}. It follows from  Eq. \eqref{dual1}  that $\widetilde{\lambda_0}\widetilde{\lambda_2}=\lambda_0\lambda_2$ and hence one may conclude that $\lambda_0\lambda_2$ is invariant under local unitary transformations. In this way, the five unitary invariants can be found to be
\begin{eqnarray}
&\widetilde{\lambda_0}\widetilde{\lambda_4}=\lambda_0\lambda_4=\sqrt{\tau}/2,&\nonumber \\
&\widetilde{\lambda_0}\widetilde{\lambda_2}=\lambda_0\lambda_2=C_{ac}/2,&\nonumber \\
&\widetilde{\lambda_0}\widetilde{\lambda_3}=\lambda_0\lambda_3=C_{ab}/2,&\\
&\left|\widetilde{\lambda_2}\widetilde{\lambda_3}-e^{i\widetilde{\varphi}}\widetilde{\lambda_1}\widetilde{\lambda_4}\right|=\left|\lambda_2\lambda_3-e^{i\varphi}\lambda_1\lambda_4\right|=C_{bc}/2,&\nonumber\\
&\widetilde{\lambda_2}\widetilde{\lambda_3}-\widetilde{\lambda_1}\widetilde{\lambda_4}\cos(\widetilde{\varphi})
=\lambda_2\lambda_3-\lambda_1\lambda_4\cos(\varphi),&\nonumber
\end{eqnarray}
where $\tau$ is three-tangle \cite{three-tangle} and $C_{ab}$ is the concurrence between qubits 1 and 2, etc.
We use Eq. \eqref{dual1} to remove the ambiguity in the definition of the canonical form and define the equivalence classes uniquely  in terms of the parameters of the canonical form. By using the property
\begin{equation}\label{kanonik_kriter}
K\equiv\frac{ \lambda _0^2 \lambda _4^2+\left|\lambda _2 \lambda _3-e^{i \varphi } \lambda _1 \lambda _4\right|{}^2}{\left(\lambda _2^2+\lambda _4^2\right) \left(\lambda _3^2+\lambda _4^2\right)}>1 (<1)
\leftrightarrow
\widetilde{K}\equiv\frac{ \widetilde{\lambda _0}^2 \widetilde{\lambda _4}^2+\left|\widetilde{\lambda _2} \widetilde{\lambda _3}-e^{i \widetilde{\varphi} } \widetilde{\lambda _1} \widetilde{\lambda _4}\right|{}^2}{\left(\widetilde{\lambda _2}^2+\widetilde{\lambda _4}^2\right) \left(\widetilde{\lambda _3}^2+\widetilde{\lambda _4}^2\right)}<1 (>1),
\end{equation}
we choose the state with $K>1$ as the canonical form. In the case where $K=1$, we obtain $\widetilde{K}=1$, $ \lambda_i=\widetilde{\lambda _i}$, and  $\widetilde{\varphi}=2\pi-\varphi$, and we choose $0\leq\varphi \leq\pi$ as the canonical state. If the three-tangle is nonzero, then the three-qubit state is of GHZ class. If three-tangle is zero and   the  reduced density matrices  $\rho_A\equiv$Tr$_{BC}\ket{\psi}\bra{\psi}$, $\rho_B$, and $\rho_C$ have rank two,
then the state $\ket{\psi}$ is a $W$-class state.

We consider that the most general local operator
\begin{equation}\label{operation}
A'=e^{i\theta_{1}}a\ket{0}\bra{0}+e^{i\theta_{2}}b\ket{0}\bra{1}+e^{i\theta_{3}}c\ket{1}\bra{0}+e^{i\theta_{4}}d\ket{1}\bra{1}\quad (a, b, c, d \  \geq0)
\end{equation}
acts on the first qubit which transforms the state given by Eq. \eqref{canonform} into
\begin{equation}
\ket{\psi'}=\frac{1}{\sqrt{p_A}}(A'\otimes I_B\otimes I_C)\ket{\psi}
\end{equation}
with a probability of $p_A=\bra{\psi}((A')^{\dag}A'\otimes I_B\otimes I_C)\ket{\psi}$ and then the resulting state can be brought into the canonical form by local unitary transformations to give
\begin{eqnarray}\label{oper1}
\ket{\psi'}&=& \frac{1}{\sqrt{p_A}}\Bigg[ \lambda_0\frac{\left|e^{i(\theta_{1}+\theta_{4})}ad-e^{i(\theta_{2}+\theta_{3})}bc\right|}{\sqrt{b^2+d^2}}\ket{000} \nonumber \\
& &+\left(\lambda_0\frac{e^{i(\theta_{1}-\theta_{2})}ab+e^{i(\theta_{3}-\theta_{4})}cd}{\sqrt{b^2+d^2}}
+e^{i\varphi}\lambda_1\sqrt{b^2+d^2}\right)\ket{100}\\
& &+\lambda_2\sqrt{b^2+d^2}\ket{101}+\lambda_3\sqrt{b^2+d^2}\ket{110}+\lambda_4\sqrt{b^2+d^2}\ket{111}\Bigg] \nonumber.
\end{eqnarray}
Using the fact that the action of the POVM operator
\begin{eqnarray}\label{operation2}
A&=&\frac{\left|e^{i(\theta_{1}+\theta_{4})}ad-e^{i(\theta_{2}+\theta_{3})}bc\right|}{\sqrt{b^2+d^2}}\ket{0}\bra{0} \nonumber \\
& &
+\left(\frac{e^{i(\theta_{1}-\theta_{2})}ab+e^{i(\theta_{3}-\theta_{4})}cd}{\sqrt{b^2+d^2}}\right)\ket{1}\bra{0} +\sqrt{b^2+d^2}\ket{1}\bra{1}
\end{eqnarray}
on the first qubit transforms the state  \eqref{canonform} into the state  \eqref{oper1} with the same probability  $p_A$, we conclude that the operators given by Eqs. \eqref{operation} and \eqref{operation2} are in the same equivalence class ($A'\equiv A$). Hence, the canonical  local operator on the first qubit can be taken as
\begin{equation}\label{canonicalA}
A=a_1\ket{0}\bra{0}+e^{i\alpha_{1}}c_1\ket{1}\bra{0}+d_1\ket{1}\bra{1}\quad (a_1, c_1, d_1\  \geq 0, \ \ 0\leq\alpha_{1}<2\pi)
\end{equation}
The fact that the probability of any transformation should be less than or equal to unity ($P=\bra{\psi}A^{\dag}A\ket{\psi}\leq 1$) implies that the eigenvalues of $A^{\dagger}A$   should be less than or equal to one. This leads to a constraint on the parameters given by
\begin{equation}\label{constraintA}
a_1^2+c_1^2+d_1^2+\sqrt{((a_1 - d_1)^2 + c_1^2) ((a_1 + d_1)^2 + c_1^2)} \leq 2.
\end{equation}
The parameters  $a_1,$ $c_1,$ and $d_1$ satisfying the inequality \eqref{constraintA} and $\alpha_{1}$ uniquely determine the equivalence class.
Now we consider that the most general transformation
\begin{equation}
B'=e^{i\beta_{1}}q\ket{0}\bra{0}+e^{i\beta_{2}}r\ket{0}\bra{1}+e^{i\beta_{3}}s\ket{1}\bra{0}+e^{i\beta_{4}}t\ket{1}\bra{1}\quad (q, r, s, t \  \geq0)
\end{equation}
is performed on the second qubit and the state given by Eq. \eqref{canonform} is transformed into the state
\begin{equation}
\ket{\psi''}=\frac{1}{\sqrt{p_B}}(I_A \otimes B'\otimes I_C)\ket{\psi}
\end{equation}
with probability $p_B= \bra{\psi}(I_A \otimes (B')^{\dag}B'\otimes I_C)\ket{\psi}$. Then the resulting state can be brought into the canonical form by local unitary transformations to give
\begin{eqnarray}\label{oper2}
\ket{\psi''}&=&\frac{1}{\sqrt{p_B}}\Bigg[ \lambda_0\sqrt{q^2+s^2}\ket{000}+\left(\lambda_1e^{i\varphi}\sqrt{q^2+s^2}+\lambda_3\frac{e^{i(\beta_{2}-\beta_{1})}qr+e^{i(\beta_{4}-\beta_{3})}st}{\sqrt{q^2+s^2}}\right)
\ket{100}\nonumber \\
& & +\left|\lambda_2\sqrt{q^2+s^2}+\lambda_4\frac{e^{i(\beta_{2}-\beta_{1})}qr+e^{i(\beta_{4}-\beta_{3})}st}{\sqrt{q^2+s^2}}\right|\ket{101}
\\
& &+\lambda_3\left|\frac{e^{i(\beta_{1}+\beta_{4})}qt-e^{i(\beta_{2}+\beta_{3})}rs}{\sqrt{q^2+s^2}}\right|\ket{110}\nonumber\\
& & +\lambda_4\left|\frac{e^{i(\beta_{1}+\beta_{4})}qt-e^{i(\beta_{2}+\beta_{3})}rs}{\sqrt{q^2+s^2}}\right|\ket{111}\Bigg].\nonumber
\end{eqnarray}
Using the result that the operator
\begin{eqnarray}
B&=&\sqrt{q^2+s^2}\ket{0}\bra{0}+
\left(\frac{e^{i(\beta_{2}-\beta_{1})}qr+e^{i(\beta_{4}-\beta_{3})}st}{\sqrt{q^2+s^2}}\right)\ket{0}\bra{1}\nonumber \\
& &+\left|\frac{e^{i(\beta_{1}+\beta_{4})}qt-e^{i(\beta_{2}+\beta_{3})}rs}{\sqrt{q^2+s^2}}\right|\ket{1}\bra{1}
\end{eqnarray}
on the second qubit transforms the state given by Eq. \eqref{canonform} into the state  given by Eq. \eqref{oper2} with the same probability $p_B$ we conclude that $B$ and $B'$ are in the same equivalence class ($B'\equiv B$). Hence, the canonical operator on the second qubit is given by
\begin{equation}\label{canonicalB}
B=a_2\ket{0}\bra{0}+e^{i\alpha_{2}}b_2\ket{0}\bra{1}+d_2\ket{1}\bra{1}\quad (a_2, b_2, d_2\  \geq0, \ \ 0\leq\alpha_{2}<2\pi).
\end{equation}
It can similarly be shown that the canonical operator on the third  qubit is of the form
\begin{equation}\label{canonicalC}
C=a_3\ket{0}\bra{0}+e^{i\alpha_{3}}b_3\ket{0}\bra{1}+d_3\ket{1}\bra{1}\quad (a_3, b_3, d_3\  \geq0, \ \ 0\leq\alpha_{3}<2\pi).
\end{equation}
The condition that eigenvalues of  $B^{\dagger}B$ and $C^{\dagger}C$ should be less than or equal to one leads to the constraint
\begin{equation}\label{constraintsBC}
a_i^2+b_i^2+d_i^2+\sqrt{((a_i - d_i)^2 + b_i^2) ((a_i + d_i)^2 + b_i^2)} \leq 2,\quad i=2, 3.
\end{equation}
The parameters  $a_i,$ $b_i$, and $d_i$ satisfying the inequality \eqref{constraintsBC} and $\alpha_{i}$ uniquely determine the equivalence classes.

\section{Optimal manipulation and  concentration of GHZ states}
In this section, we discuss the optimal manipulation of three-qubit pure states using \emph{one successful branch protocol} (OSBP). In this protocol, we maximize local success probabilities and the state is either transformed into the desired one or else the particle is disentangled from other particles. Maximization of the probability imposes some conditions on the operators. Two operators $A$ and $\tilde{A}=\alpha A$ ($\alpha$: constant) make the same transformations on states
\begin{equation}
\ket{\psi'}=\frac{1}{\sqrt{p}}A\ket{\psi},\quad \ket{\psi'}=\frac{1}{\sqrt{\tilde{p}}}\tilde{A}\ket{\psi}
\end{equation}
with probabilities $p$ and  $\tilde{p}=|\alpha| p$, respectively. By choosing $\alpha$ it is possible to increase or decrease the transformation probability for any transformation. However, the value of $\alpha$ and hence the probability is restricted by the condition that the greater eigenvalue of $A^{\dag}A$ can not exceed one. Hence, the transformation probability is maximum when the greater eigenvalue is one [i. e., $\textnormal{det}(I_A -A^{\dag} A)=0$]. In this case, for a two-outcome POVM with elements $A$ and $\bar{A}$, satisfying $\bar{A}^{\dag}\bar{A}+A^{\dag} A=I_A$, the rank of $\bar{A}$ is one, and hence   the state $(\bar{A}\otimes I_B \otimes I_C)\ket{\psi}$ is a product state in the first particle.  Similar considerations can be done for the measurement of the second and third particles with the  POVM  elements $B,\ \bar{B}$ and $C,\ \bar{C}$ satisfying $\bar{B}^{\dag}\bar{B}+B^{\dag} B=I_B$ and $\bar{C}^{\dag}\bar{C}+C^{\dag} C=I_C$. If the transformation $A\otimes B \otimes C\ket{\psi}$ gives the desired state then the probability maximization of this transformation requires
\begin{equation}\label{constraints3}
\textnormal{det}(I_A -A^{\dag} A)=0,\ \textnormal{det}(I_B-B^{\dag} B)=0,\ \textnormal{det}(I_C -C^{\dag} C)=0
\end{equation}
which imply that the state is either transformed into the  desired state  or otherwise disentangled (i.e., we are using  OSBP).

The problem is the optimal transformation of the generic state given by Eq. \eqref{canonform}  into the state
\begin{equation}\label{canonform2}
\ket{\psi'}=\lambda^{'}_0\ket{000}+\lambda^{'}_1e^{i\varphi'}\ket{100}+\lambda^{'}_2\ket{101}+\lambda^{'}_3\ket{110}+\lambda^{'}_4\ket{111},
\quad \lambda^{'}_i \geq 0.
\end{equation}
by the action of all three parties
\begin{eqnarray}\label{transformed}
\ket{\psi'}&=&\frac{1}{\sqrt{P}}A\otimes B \otimes C\ket{\psi}\nonumber \\
&=&\frac{1}{\sqrt{P}}\Big(\lambda_0a_1a_2a_3\ket{000} \\
&+&\big((\lambda_0c_1e^{i\alpha_{1}}+\lambda_1e^{i\varphi }d_1)a_2a_3+\lambda_2d_1a_2b_3e^{i\alpha_{3}}
+\lambda_3d_1b_2e^{i\alpha_{2}}a_3+\lambda_4d_1b_2b_3e^{i(\alpha_{2}+\alpha_{3})}\big)\ket{100}\nonumber \\ &+&\left|(\lambda_2d_1a_2d_3+\lambda_4d_1b_2e^{i\alpha_{2}}d_3)\right|\ket{101}+\left|(\lambda_3d_1d_2a_3+\lambda_4d_1d_2b_3e^{i\alpha_{3}})\right|\ket{110}+\lambda_4d_1d_2d_3\ket{111}\Big)\nonumber
\end{eqnarray}
where $P=\bra{\psi}A^{\dag}A\otimes B^{\dag}B\otimes C^{\dag}C\ket{\psi}$. We now impose that the transformed state given by Eq. \eqref{transformed} is the GHZ state and obtain
the conditions
\begin{equation}\label{constraints}
c_1e^{i\alpha_{1}}=\frac{\lambda _2 \lambda _3-e^{i\varphi } \lambda _1 \lambda _4}{\lambda_{0}\lambda_{4}}d_1,\ b_2=\frac{\lambda _2}{\lambda _4}a_2,\ b_3=\frac{\lambda _3}{\lambda _4}a_3,\ \alpha_{2}=\alpha_{3}=\pi, \ \lambda_0a_1a_2a_3=\lambda_4d_1d_2d_3
\end{equation}
and the probability of success  turns out to be
\begin{equation}\label{prob}
P=2a_1^2a_2^2a_3^2\lambda_{0}^2.
\end{equation}
The maximization of the local probabilities given by Eq. \eqref{constraints3} leads to the following constraints
\begin{eqnarray}\label{cond}
(1-a_1^2)(1-d_1^2)&=& c_1^2=\frac{C_{bc}^2}{\tau}d_1^2,\nonumber \\
(1-a_2^2)(1-d_2^2) &=& b_2^2=\frac{C_{ac}^2}{\tau}a_2^2, \\
(1-a_3^2)(1-d_3^2)&=& b_3^2=\frac{C_{ab}^2}{\tau}a_3^2.\nonumber
\end{eqnarray}
The problem of  optimal concentration of the GHZ state using OSBP is reduced to the problem of the maximization of the probability given by Eq. \eqref{prob} subject to the constraints given by Eqs. \eqref{constraintA}, \eqref{constraintsBC}, \eqref{constraints} and \eqref{cond}. The solution for the most general case, where all bipartite entanglements  $C_{ab}$, $C_{ac}$, and $C_{bc}$ are nonzero, requires numerical calculations. However we find  the analytical solutions in terms of the concurrences and the three tangle in the cases where at least one of the concurrences is zero. In these cases, complex phases in the states  can be eliminated using local unitary transformations. For example, in the case   $C_{ab}=0$ ($\lambda _3=0$)  the general state turns out to be
\begin{equation}\label{canform7}
\ket{\psi}=\lambda_0\ket{000}+e^{i\varphi}\lambda_1\ket{100}+\lambda_2\ket{101}+\lambda_4\ket{111}.
\end{equation}
However, the complex phase can be eliminated by the local unitary transformations
\begin{equation}\label{uniter8}
U=(\ket{0}\bra{0}+e^{-i\varphi}\ket{1}\bra{1})\otimes I_B \otimes (\ket{0}\bra{0}+e^{i\varphi}\ket{1}\bra{1}).
\end{equation}

i) States with no bipartite entanglement ($C_{ab}=C_{ac}=C_{bc}=0$): In this case the canonical form of the state is given by
\begin{equation}\label{canform1}
\ket{\psi}=\lambda_0\ket{000}+\lambda_4\ket{111}
\quad (\lambda_0 \geq \lambda_4).
\end{equation}
The local operation $\frac{\lambda_4}{\lambda_0}\ket{0}\bra{0}+\ket{1}\bra{1}$ on one of the qubits is sufficient to obtain the GHZ states with the maximum success probability given by
\begin{equation}\label{prob1}
P_{max}=1-\sqrt{1-\tau}.
\end{equation}
Since the three tangle, $\tau$, is an entanglement monotone, so is $P_{max}$.

ii) States with only one nonzero bipartite entanglement (e.g., $C_{ab}=C_{ac}=0,\ C_{bc}\neq0$): In this case the canonical form of the state is given by
\begin{equation}\label{canform2}
\ket{\psi}=\lambda_0\ket{000}+\lambda_1\ket{100}+\lambda_4\ket{111}.
\end{equation}
We find that the concentration of the GHZ state depends solely on the party A: The parties B and/or C  cannot obtain the GHZ state, but only party A has this privilege.  The solutions for the local operators and the maximum probability are found to be
\begin{equation}\label{prob2}
P_{max}=1-\sqrt{1-\tau},
\end{equation}
\begin{equation}\label{soloper2}
A=\frac{\sqrt{P_{max}}}{\sqrt{2}\lambda_0}\ket{0}\bra{0}-
\frac{\lambda_1 \sqrt{P_{max}}}{\sqrt{2}\lambda_0 \lambda_4}\ket{1}\bra{0}+
\frac{\sqrt{P_{max}}}{\sqrt{2}\lambda_4}\ket{1}\bra{1},\ B=I,\ C=I.
\end{equation}
We note that if the bipartite entanglement exists only between two qubits, the concentration probability depends only on the
three-tangle, not on the bipartite concurrence as given by Eq. \eqref{prob2}. From symmetry it is straightforward to find the optimal distillation protocols for the other states in this class: The states
\begin{eqnarray}\label{canform11}
\ket{\psi}&=&\lambda_0\ket{000}+\lambda_2\ket{101}+\lambda_4\ket{111},\nonumber \\
\ket{\psi}&=&\lambda_0\ket{000}+\lambda_3\ket{100}+\lambda_4\ket{111}
\end{eqnarray}
can be optimally transformed to the GHZ state by the operators
\begin{eqnarray}\label{soloper11}
A&=&I,\ B=\frac{\sqrt{P_{max}}}{\sqrt{2}\lambda_0}\ket{0}\bra{0}-
\frac{\lambda_2 \sqrt{P_{max}}}{\sqrt{2}\lambda_0 \lambda_4}\ket{0}\bra{1}+
\frac{\sqrt{P_{max}}}{\sqrt{2}\lambda_4}\ket{1}\bra{1},\  C=I, \nonumber \\
A&=&I,\ B=I,\ C=\frac{\sqrt{P_{max}}}{\sqrt{2}\lambda_0}\ket{0}\bra{0}-
\frac{\lambda_3 \sqrt{P_{max}}}{\sqrt{2}\lambda_0 \lambda_4}\ket{0}\bra{1}+
\frac{\sqrt{P_{max}}}{\sqrt{2}\lambda_4}\ket{1}\bra{1},
\end{eqnarray}
respectively, with the maximum probability given by Eq. \eqref{prob2}.

(iii) States with only one vanishing bipartite entanglement (e.g., $C_{ab}=0,\ C_{ac}\neq 0,\ C_{bc}\neq 0$): In this case, the canonical form of the state is given by
\begin{equation}\label{canform3}
\ket{\psi}=\lambda_0\ket{000}+\lambda_1\ket{100}+\lambda_2\ket{101}+\lambda_4\ket{111}.
\end{equation}
We find that the condition  $C_{ab}=0$ together with the maximization of the local probabilities given by Eq. \eqref{cond} leads us to the result that $a_3=d_3=1,\ b_3=0$, and any operation other than the unitary transformation on the third qubit reduces the probability of obtaining the GHZ state. By using the transformation given by Eq. \eqref{transformed}, it can be shown that neither  party A nor  party B  can concentrate the GHZ state alone, but the combined action of both parties A and B is necessary to obtain the GHZ state. We find the solution for the maximum success probability given by
\begin{equation}\label{prob3}
P_{max}=1+\frac{C_{ac}C_{bc}}{\sqrt{\tau}}-\sqrt{\left(1+\frac{C_{ac} C_{bc}}{\sqrt{\tau}}\right)^2-\tau}.
\end{equation}
To prove that no concentration protocol can give a greater probability, one needs to show that the inequality
\begin{equation}\label{mono1}
P(\ket{\psi})\geq\sum_{i}p_iP(\ket{\psi_i})
\end{equation}
is satisfied for any sequence of local quantum operations that transform $\ket{\psi}$ into $\ket{\psi_i}$ with a probability $p_i$. The right-hand side of the inequality \eqref{mono1} is the average probability of obtaining the GHZ state  using several branches, whereas the left-hand side is the probability for OSBP. By taking into account that any POVM can be decomposed into a sequence of two-outcome POVMs \cite{Optimal_GHZ_distil}, it is sufficient to show
\begin{equation}\label{mono2}
P(\ket{\psi})\geq p_1P(\ket{\psi_1})+p_2P(\ket{\psi_2}),
\end{equation}
where $\ket{\psi_1}$ and $\ket{\psi_2}$ are obtained by the most general POVMs on one of the qubits. We start with a two outcome POVM with operators
\begin{eqnarray}\label{oper7}
A_1&=&a_1\ket{0}\bra{0}+c_1e^{i\alpha_{1}}\ket{1}\bra{0}+d_1\ket{1}\bra{1},\nonumber \\   A_2&=&f_1\ket{0}\bra{0}+g_1e^{i\beta_{1}}\ket{1}\bra{0}+h_1\ket{1}\bra{1}
\end{eqnarray}
acting on the first qubit and satisfying $A_1^{\dag}A_1+A_2^{\dag}A_2=I$. The states
\begin{eqnarray}\label{stat3}
\ket{\psi_1} &=& \frac{1}{\sqrt{p_1}}(\lambda_0a_1\ket{000}+(\lambda_0c_1e^{i\alpha_{1}}+\lambda_1d_1)\ket{100}\nonumber \\ & & +\lambda_2d_1\ket{101}+\lambda_4d_1\ket{111}),\nonumber \\
\ket{\psi_2} &=& \frac{1}{\sqrt{p_2}}(\lambda_0f_1\ket{000}+(\lambda_0g_1e^{i\beta{1}}+\lambda_1h_1)\ket{100} \\ & & +\lambda_2h_1\ket{101}+\lambda_4h_1\ket{111})\nonumber
\end{eqnarray}
are obtained with probabilities $p_i=\bra{\psi}A_i^{\dag}A_i\otimes I_B \otimes I_C\ket{\psi}$. The complex phases in Eq. \eqref{stat3} can be eliminated by local unitary transformations and then  OSBP is used on the states $\ket{\psi_1}$ and $\ket{\psi_2}$ to give the maximum probabilities  $P(\ket{\psi_1})$ and $P(\ket{\psi_2})$ for the concentration  of the GHZ state.  To check if the inequality (\ref{mono2}) is satisfied, we maximize   $p_1P(\ket{\psi_1})+p_2P(\ket{\psi_2})$ and find that the maximum is obtained for  $P(\ket{\psi_1})=0$ or $P(\ket{\psi_2})=0$, which proves that  no concentration protocol can produce a higher probability of success than the OSBP  we present.

\section{Results and discussion}

One of the main difficulties in the optimal manipulation problem of multipartite entangled states is  that there are too many parameters coming from the general forms of the states and local operations. The use of equivalence classes significantly reduces the number of parameters as the states that can be transformed into each other by local unitary transformations are equal from the information theoretic point of view. In addition to the equivalence classes of states, we define the equivalence classes of local measurements such that local operations which transform states from one equivalence class into another with the same probability are equivalent. This approach does not only simplify the concentration problem, but also the results arise in terms of the local unitary invariants of the quantum states, namely, bipartite concurrences and the three-tangle.

We find that when there is no bipartite entanglement between particles ($C_{ab}=C_{ac}=C_{bc}=0$), the optimal concentration of the GHZ state can be obtained by a single measurement on any one of the particles. In the case where there is bipartite entanglement between only two particles (e.g., $C_{ab}=C_{ac}=0,\ C_{bc}\neq0$), the optimal concentration of the GHZ state can only be done by a measurement on the particle which has no bipartite entanglement with the other two as given by Eq. \eqref{soloper2}.
We obtain the interesting result that the maximum success probabilities for both cases, states with no bipartite entanglement and states with only one nonzero bipartite entanglement, depend only on the three-tangle as given by Eqs. \eqref{prob1} and \eqref{prob2}.
We also find that the optimal concentration of states  with only one vanishing bipartite entanglement (e.g., $C_{ab}=0,\ C_{ac}\neq 0,\ C_{bc}\neq 0$) can only be obtained by local measurements performed by the parties $A$ and $B$ with the maximum success probability given by Eq. \eqref{prob3}.
We may conclude that the use of equivalence classes of local measurement operators simplifies the optimal manipulation problem and may be used for the manipulation of other multi-partite states.

\section{ACKNOWLEDGMENT}

This work has been partially supported by the Scientific and
Technological Research Council of Turkey (TUBITAK) under Grant 113F256.

\end{document}